\documentclass[12pt]{article}

\usepackage{amsmath}
\usepackage{graphicx}
\usepackage{amssymb}
\usepackage{amsfonts}
\usepackage{latexsym}

\def\beq{\begin{eqnarray}}
\def\eeq{\end{eqnarray}}

\renewcommand{\vec}[1]{{\bf #1}}

\begin{document}
\begin{titlepage}
\title{Bouncing Universes in Scalar-Tensor Gravity Models admitting Negative Potentials}
\author{B. Boisseau$^1$\thanks{email:bruno.boisseau@lmpt.univ-tours.fr}, 
H. Giacomini$^1$\thanks{email:hector.giacomini@lmpt.univ-tours.fr},
D. Polarski$^2$\thanks{email:david.polarski@umontpellier.fr},
A. A. Starobinsky$^{3,4}$\thanks{email:alstar@landau.ac.ru}\\
\hfill\\
$^1$Universit\'e de Tours, Laboratoire de Math\'ematiques et Physique Th\'eorique,\\
CNRS/UMR 7350, 37200 Tours, France\\
$^2$Universit\'e Montpellier \& CNRS, Laboratoire Charles Coulomb,\\
UMR 5221, F-34095 Montpellier, France\\
$^3$L. D. Landau Institute for Theoretical Physics RAS,\\
 Moscow 19334, Russian Federation\\
$^4$Kazan Federal University, Kazan 420008, \\
Republic of Tatarstan, Russian Federation}

\pagestyle{plain}
\date{\today}

\maketitle

\begin{abstract}
We consider the possibility to produce a bouncing universe in the framework of 
scalar-tensor gravity models in which the scalar field potential may be negative, 
and even unbounded from below. 
We find a set of viable solutions with nonzero measure in the space of initial 
conditions passing a bounce, even in the presence of a radiation component, 
and approaching a constant gravitational coupling afterwards. Hence we have a model 
with a minimal modification of gravity in order to produce a bounce in the early universe 
with gravity tending dynamically to general relativity (GR) after the bounce. 
\end{abstract}
\end{titlepage}

\section{Introduction}

Since the realization that the strong curvature singularity arising in the past of our 
Universe in Friedmann-Lema\^itre-Robertson-Walker (FLRW) models may not be avoided in 
generic solutions of the Einstein gravity with matter in the form of an ideal fluid 
(only its type changes from an isotropic one to the generic BKL vacuum singularity), 
there were many attempts to construct solutions avoiding this singularity by having at 
least one bounce at some high value of curvature either using more complicated 
field-theoretical models of matter or by modifying gravity. The simplest example to 
obtain one bounce in a set of solutions having a nonzero measure in the space of initial 
conditions 
is presented by a massive scalar field in a closed (i.e. positively spatially curved) FLRW 
universe~\cite{Star78}. This does not remove the curvature singularity in a generic 
solution, even in the class of FLRW models, but shifts it to the past, beyond any given 
finite number of bounces of the scale factor. Non-singular FLRW solutions with an infinite 
number of bounces exist, too, but they are degenerate: they have a zero measure in the 
space of initial conditions~\cite{Page84}, see also \cite{Ka98}. This occurs because a 
minimally coupled scalar field with a non-negative potential can violate the strong energy 
condition, but not the 
weak and null ones.\footnote{In the case of open (negatively spatially curved) FLRW models, 
a bouncing solution was found in~\cite{MO79} for a non-minimally coupled scalar field with 
a quartic potential  (so this model represents a particular kind of scalar-tensor gravity 
models). However, later it was shown in~\cite{Star81} that this regular solution is 
degenerate, too: it is unstable against arbitrarily small anisotropic and/or inhomogeneous 
metric perturbations preventing transition to the regime with a tensor ghost (repulsive 
gravity).} 
However, having a bounce due to a positive spatial curvature requires severe fine tuning of 
initial conditions before the contraction stage in the past, see e.g. the calculations of the 
measure of those ones leading to a bounce in the above mentioned cosmological model 
in~\cite{Star78} and ~\cite{GT08}. That is why it is much more interesting to obtain a 
non-degenerate 
bounce in FLRW models in the absence of spatial curvature. Previously known examples of such 
FLRW models are based on such radical modifications of general relativity (GR) as theories with
scalar~\cite{DSNA11,BS13} or tensor ghosts, loop quantum 
gravity (see e.g. \cite{APS06}) or gravity described by an effectively non-local 
Lagrangian~(see e.g.\cite{QECZ11},\cite{ESV11}, and \cite{BP14} for a recent review).
By contrast, in our paper we would like to restrict ourselves to the well-known and very modest 
modification of GR - scalar-tensor gravity without ghosts. On the other hand, we want to 
abandon the common assumption that the scalar field potential $V(\phi)$ in this gravity 
theory is non-negative, we will even consider the case when it is unbounded from below. 
In Section 2, the bouncing model is presented and general equations and results are given. 
In Section 3, the existence of physically viable bouncing solutions is shown and their 
behaviour is studied in details. Section 4 contains our conclusions. 
\section{A bouncing model}

Let us consider a universe where gravity is described by scalar-tensor theory. The 
Lagrangian density in the Jordan frame of the gravitational sector is given by 
\beq
\label{L}
L = \frac{1}{2}\left[ F(\Phi)R - Z(\Phi)~g^{\mu\nu}\partial{\mu}\Phi\partial{\nu}\Phi 
                                              - 2 U(\Phi) \right]~.
\eeq
Everywhere below we will use the freedom to take $Z=1$ or $Z=-1$, corresponding 
physically to $\omega_{BD}>0$ or $\omega_{BD}<0$ ($\omega_{BD} = 
Z F~\left(\frac{dF}{d\Phi}\right)^{-2}$). 
For $\omega_{BD}<0$, the theory is ghost-free provided $-\frac{3}{2}<\omega_{BD}<0$.  
We consider further spatially flat FLRW universes with metric $ds^2=-dt^2+a^2(t)d\vec{x}^2$ 
yielding the following modified Friedmann equations 
\beq
-3 F H^2 + \frac12 Z~\dot{\Phi}^2 - 3 H \dot{F} + U &=& 0~, \label{Fr1}\\
2 F \dot{H} + Z~\dot{\Phi}^2 + \ddot{F} - H \dot{F} &=& 0~. \label{Fr2}
\eeq
with $H\equiv \frac{\dot{a}}{a}$. Here and below a dot, resp. a prime, stands for the 
derivative with respect to $t$, resp. to $\Phi$. 
Equations \eqref{Fr1},\eqref{Fr2} contain the equation of motion of $\Phi$
\beq
Z~(-\ddot{\Phi} - 3 H \dot{\Phi}) + 3 F'(\dot{H}+2 H^2) - U' = 0~. \label{ddPhi}
\eeq
We start our analysis with $Z=1$ and we specialize to the following model
\beq
F &=& -\frac{1}{6} \Phi^2 + \kappa^{-2}, \label{F}\\
U &=& \frac{\Lambda}{\kappa^2} - c \Phi^4 ~,\label{U}
\eeq
where $\kappa^{-2}>0$, $\Lambda>0$ and $c$ are constant parameters, only 
$c$ being dimensionless. Surprisingly, while we arrived at this model by requiring that 
a combination of \eqref{Fr1}, \eqref{Fr2} and \eqref{ddPhi}, without derivation with 
respect to $t$, be zero, this model has been considered in the past in completely 
different contexts (see e.g. \cite{MTZ03}).
Thus \eqref{ddPhi} corresponds to a massless scalar field conformally coupled to Einstein 
gravity with an interaction potential $-c\Phi^4$.
Therefore, the energy-momentum tensor of the field $\Phi$ is the sum of ``dark'' 
radiation with a negative energy density and a cosmological constant.

As a result, the system \eqref{Fr1}, \eqref{Fr2} reduces to
\beq
3H^2 = \Lambda + \kappa^2 \frac{A}{a^4}, \label{Eq1}\\
\frac{1}{2}\left(\frac{d\chi}{d\eta}\right)^2 - c\chi^4 = A~, \label{Eq2}
\eeq
with an effective positive cosmological constant $\Lambda$ and the bare gravitational 
constant $8\pi G = F^{-1}(\Phi=0)\equiv \kappa^2$, while $\chi = a \Phi$ and 
$\eta=\int dt/a(t)$. 
The crucial point is that the constant $A$ -- the energy density of the 
field $\chi$ in Minkowski space-time -- can be negative here.

For $A<0$, we have in front of us an explicit illustration of the possibility for 
scalar-tensor models to accommodate an effective dark energy component of the 
phantom type ($w_{eff}<-1$) \cite{BEPS00},\cite{GPRS06} \footnote{A phenomenological dark energy 
model in GR based on \eqref{Eq1} was considered in \cite{P13}, leading to \eqref{a} and 
\eqref{H}.}. 
Our system can be completely integrated. As we are interested in bouncing solutions, we must 
take $A<0$ and from \eqref{Eq2}, we have $c>0$ so that $U$ is necessarily an \emph{inverted} 
potential, unbounded from below. In spite of looking unphysical at first sight, scalar fields 
with such an interaction have been often considered both in quantum field theory and 
cosmology, see e.g.~\cite{Rub09}. 

Integrating \eqref{Eq1}, a bouncing solution is obtained 
\beq
a &=& a_0 \cosh^{\frac{1}{2}}\Big[2 \sqrt{ \frac{\Lambda}{3} }~t\Big]~, \label{a}\\
H &=& \sqrt{ \frac{\Lambda}{3} } \tanh \Big[2 \sqrt{\frac{\Lambda}{3}}~t\Big]~,\label{H}
\eeq
where $a_0=\left(\frac{-A \kappa^2}{\Lambda}\right)^{\frac{1}{4}}$ is the value of $a$ at the 
bounce located at $t=0$ with a trivial redefinition of $t$. 
It satisfies $\dot{H} > 0$ and has a constant Ricci scalar $R=6(\dot{H}+2H^2)= 4 \Lambda$.
Integrating \eqref{Eq2} the analytical expression for $\chi$ is obtained in terms of the 
Jacobi elliptic function ${\rm dn}(u|m)$ \cite{AbrSte} as follows  
\beq
\chi(\eta) = -\left( \frac{-A}{c}\right)^{\frac{1}{4}} \frac{1}{{\rm dn}(u|2)}  ~,\label{chi}
\eeq
with
\beq
u = {\rm dn}^{-1} \left( -\left( \frac{\Lambda}{c \kappa^2}\right)^{\frac{1}{4}} 
                                              \frac{1}{\Phi_0}\,|\,2 \right)              
                 + \sqrt{2} \left( -Ac \right)^{\frac{1}{4}} ~\eta ~,  \label{u}
\eeq 
 where ${\rm dn}^{-1}(u|m)$ is the inverse function of ${\rm dn}(u|m)$ and $\Phi_0=\Phi(t=0)$. 
The conformal time $\eta(t)$, chosen here such that $\eta(0) = 0$, is given by   
\beq
\eta(t) =- \frac{i}{a_0 \sqrt{\frac{\Lambda}{3}}}F\left(i\sqrt{\frac{\Lambda}{3}}~t~|~2\right) 
                                       ~,\label{eta}
\eeq
where $F(x|m)$ is the elliptic integral of the first kind \cite{AbrSte}. This function is 
odd and then $\eta(t)$ is real. It is possible to show that
\beq
\eta(\infty) = \frac{1}{a_0 \sqrt{\frac{2\Lambda}{3}}} 
              \frac{\Gamma^2\left(\frac{1}{4}\right)}{4\sqrt{\pi}} ~. \label{etainfty}
\eeq
So we get finally 
\beq
\Phi(t) = \frac{ -\left( \frac{\Lambda}{c \kappa^2}\right)^{\frac{1}{4}} }
{\sqrt{\cosh(2 \sqrt{\frac{\Lambda}{3}}~t)}}  \frac{1}{{\rm dn}\left({\rm dn}^{-1}
\left(-\left( \frac{\Lambda}{c \kappa^2}\right)^{\frac{1}{4}} \frac{1}{\Phi_0}~|~2 \right)  
-i\sqrt{2}(\frac{9c}{\kappa^2\Lambda})^{\frac{1}{4}}
                F\left(i\sqrt{\frac{\Lambda}{3}}~t~|~2\right)~|~2\right) }~.\label{Phit}
\eeq
We note that under the transformation $Z\to -1,~F\to -F~,U\to -U$, the same solution 
\eqref{Phit} is obviously obtained and that \eqref{Eq1} and \eqref{Eq2} remain valid 
as well because \eqref{Fr1},\eqref{Fr2} and \eqref{ddPhi} are invariant under this
transformation. 
 
Another attractive feature of this model is the possibility to have a 
bounce in the presence of a true radiation component resulting in a 
trivial change $A\to A'$ in \eqref{Eq1}. This can model a bounce in the primordial universe. 
The bounce will survive provided the sum of ``dark'' and true radiation remains 
negative or
\beq
A + \rho_{rad,0} ~a'^4_0 \equiv A' = - \frac{\Lambda}{\kappa^2} ~a'^4_0 <0~, \label{A'}
\eeq
which shows that $\frac{A}{a^4_0} = \frac{A'}{a'^4_0}$ is invariant, the scale factor $a'_0$  
at the bounce decreases in the presence of radiation and also when $A$ increases 
towards zero. Let us finally note that the bounce disappears for $A'\ge 0$. 
%

\section{Study of the bouncing solution}

In spite of the fact that we have obtained an explicit closed form expression for 
$\Phi(t)$ we will study the behaviour of the bouncing solution by applying the 
qualitative theory of differential equations. 
This will provide us with a more intuitive and direct approach and will enable us 
to obtain results about the qualitative behaviour of our system otherwise 
difficult to find from \eqref{Phit}.  
Solving for the wave equation recast into the form
\beq
\ddot{\Phi} + 3 H \dot{\Phi} + 4 c \Phi ( \tilde{\Phi}^2 - \Phi^2 ) = 0~,\label{ddPhi2}
\eeq
where $\tilde{\Phi}\equiv (\frac{\Lambda}{6c})^{\frac{1}{2}}$, and evaluating \eqref{Fr1} 
at the bounce 
\beq
\frac12 \dot{\Phi}_0^2  - c ( \Phi_0^4 - \Phi_{0,min}^4 ) = 0~, \label{ic}
\eeq
where $\Phi_{0,min}\equiv (\frac{\Lambda}{\kappa^2 c})^{\frac{1}{4}} = 
(\frac{-A}{c a^4_0})^{\frac{1}{4}}$, our system is automatically solved.

The inequality $\Phi^2 < 6 \kappa^{-2}$ is readily obtained from the requirement 
$F>0$ while $\Phi=0$ is impossible from \eqref{Eq2}.
Choosing $\Phi_0 > 0$, the following inequality is further obtained
\beq
\tilde{\Phi} < \Phi_{0,min} \leq \Phi_0 <  \Phi_{max}~, \label{int2}
\eeq
with 6 $\kappa^{-2} \equiv \Phi^2_{max}$.
Indeed, from \eqref{Fr1}, the condition $U(\Phi_0)\equiv U_0\le 0$ implies immediately 
$\Phi_0\ge \Phi_{0,min}$, while $\Phi < \Phi_{max}$, valid at all times, 
follows from the physical condition $F > 0$. 
A non vanishing interval allowed for $\Phi_0$ requires 
$\Phi_{0,min} < \Phi_{max}$ or 
\beq
\frac{\Lambda~\kappa^2}{36c} < 1~, \label{parcond}
\eeq
implying also $\tilde{\Phi}<\Phi_{0,min}$.
From \eqref{ic}, with $\dot{\Phi}_0 < 0$ (the justification of this choice 
will be clear below), we obtain 
\beq
\dot{\Phi}_0 = -\sqrt{ 2c( \Phi_0^4 - \Phi_{0,min}^4 ) },   \label{dPhi0}
\eeq
with $\dot{\Phi}_0$ decreasing when $\Phi_0$ is increasing. 

\begin{figure}
\begin{centering}
\includegraphics[scale=.8]{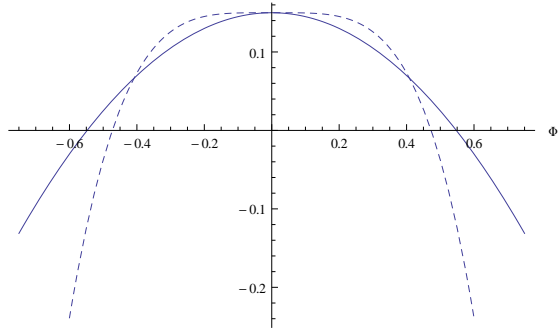}
\par\end{centering}
\caption{The functions $F$ (full) and $U$ (dashed) are displayed for 
the model parameters $\kappa^2 = 20$, $c = 3$ and $\Lambda = 3$. The two curves coincide 
at $\Phi=0$ because we have actually plotted $\Lambda F$. While $U$ 
vanishes at $\Phi\equiv \Phi_{0,min}=(\frac{\Lambda}{\kappa^2 c})^{\frac{1}{4}}$, $F$ does so at 
$\Phi\equiv \Phi_{max}=\frac{\sqrt{6}}{\kappa}$. For these parameters we have 
$\Phi_{0,min}=0.4728,~\Phi_{0,cr} = 0.4936860,~\Phi_{max}= 0.5477$. For $\Phi>0$, the range 
of possible values at the bounce lie in the interval $\Phi_{0,min}\le \Phi < \Phi_{max}$ but 
only values in the interval $\Phi_{0,cr}\le \Phi < \Phi_{max}$ and with a negative 
slope will yield physically viable solutions.}
\label{fig1}
\end{figure}

We study the dynamics for $t\geq 0$. For $t<0$ the dynamics 
follows immediately due to the invariance of the system under reflection  
$t\to -t$. We will come back to this point at the end of our analysis. We will 
analyse now the possible existence of local maxima and minima for $\Phi(t)$.
From \eqref{Eq2} we readily obtain for $\dot{\Phi}=0$
\beq
\Phi^2 = \frac{ \frac{1}{2} H^2 +\sqrt{ \frac{H^4}{4} - 4 c A a^{-4} } }{2c}~.\label{Phi2}
\eeq
The choice of the positive sign before the square root is compulsory since $\Phi^2$ 
must be positive. It is straightforward to show that \eqref{parcond} implies 
$\Phi<\Phi_{max}$ for $\Phi$ given by (\ref{Phi2}). 
Hence the inequality $\Phi<\Phi_{max}$ follows when $\dot{\Phi}=0$. We have further 
from \eqref{Fr2} when $\dot{\Phi}=0$ that $-\Phi \ddot{\Phi} + 6 F \dot{H}=0$, 
therefore $\ddot{\Phi}>0$. So we have shown the impossibility for $\Phi$ to have a maximum.  

On the other hand if $\Phi$ has a finite limit for $t \rightarrow \infty$, we 
must have $\dot{\Phi}_\infty=0 $ and $\ddot{\Phi}_\infty=0 $, therefore \eqref{ddPhi2} gives
$4 c \Phi_\infty ( \tilde{\Phi}^2 - \Phi_\infty^2 ) = 0$.
Hence we have either $\Phi_\infty = 0$ or $\Phi_\infty = \tilde{\Phi}$.
These results imply in particular that had we chosen $\dot{\Phi}_0 > 0$, $\Phi$ would 
eventually enter the unphysical domain $\Phi \ge \Phi_{max}$, so this 
choice is not allowed.  

\begin{figure}
\begin{centering}
\includegraphics[scale=.7]{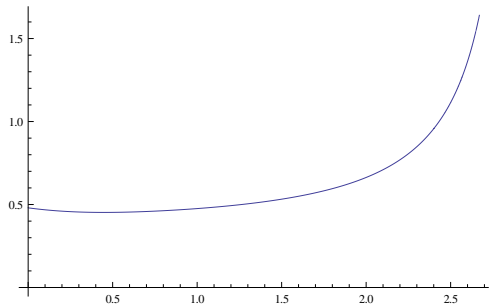}~~
\par\end{centering}
\caption{The function $\Phi(t)$ is shown starting at $t=0$ at the bounce for the same model 
parameters as in Figure 1. The initial value $\Phi_0$ is slightly below the critical value 
$\Phi_{0,cr} = 0.4936860...$ so this solution will diverge in a finite time.} 
\label{fig2}
\end{figure}

If the solution has a minimum, it is unique as $\Phi$ does not have any maximum. 
This minimum must be larger that $\tilde{\Phi}$ as a direct consequence of (\ref{ddPhi2}). 
Hence after the minimum, $\dot{\Phi} > 0$ forever since there is no maximum and then $\Phi$ 
goes to infinity -- therefore crossing the value $\Phi_{max}$ -- because the only possible 
finite limits $\Phi_\infty$ are lower than the minimum and cannot be reached. We will now 
show that in this case $\Phi$ tends to infinity in a finite time $t_{\infty}$. 
When $\Phi \rightarrow \infty$, keeping the leading terms in \eqref{Fr1} we easily find 
$\dot{\Phi} \sim \sqrt{2 c}\Phi^2$, and after integration 
\beq
\Phi \sim \frac{1}{-\sqrt{2 c}t + k} .
\eeq
with $k > 0$ and $t_{\infty} = \frac{k}{\sqrt{2 c}}$.
As a consequence $F$ becomes zero in a finite time too, rendering this 
solution unphysical. When $\Phi_0=\Phi_{0,min}$ $(U_0=0)$, the bounce is a 
minimum from eqs.(\ref{ddPhi2}) and then $\Phi$ goes to infinity in a 
finite time (see Figure \ref{fig2}).

Let us consider now the case where there is no minimum, i.e. $\dot{\Phi}<0$ for $t\ge 0$. 
From \eqref{ddPhi2} we get $\ddot{\Phi} = -3 H \dot{\Phi} - 
4c\Phi (\tilde{\Phi}^2-\Phi^2) > 0$ as long as $\Phi > \tilde{\Phi}$.
However when $\Phi(t) < \tilde{\Phi}$, the sign of $\ddot{\Phi}(t)$ can change.
Two subcases will therefore arise:

1) $\Phi(t)$ is always greater than $\tilde{\Phi}$ and, since $\dot{\Phi}(t) < 0$, 
tends asymptotically to $\Phi=\tilde{\Phi}$ with positive concavity ($\ddot{\Phi}(t)>0$).

2) $\Phi(t)$ crosses $\tilde{\Phi}$ and, since $\dot{\Phi}(t)<0$, tends asymptotically to $0$.
Below $\tilde{\Phi}$, $\ddot{\Phi}(t)$ can however change sign.

In order to differentiate these two possibilities, we study the system in the neighborhood 
of $t=\infty$ where (\ref{ddPhi2}) can be recast as an autonomous planar system, viz.
\beq
\dot{y_1} &=& y_2, \label{AS1}  \\  
\dot{y_2} &=& - \sqrt{3 \Lambda} ~y_2 - 2 \frac{\Lambda}{3} ~y_1 + 4c y_1^3~, \label{AS2}
\eeq
with $y_1\equiv \Phi$.  
This system has two hyperbolic fixed (critical) points: $(0,0)$ and $(\tilde{\Phi},0)$.
The eigenvalues associated to $(0,0)$ are both negative, $\lambda_1 = 
-\sqrt{\frac{\Lambda}{3}}$ and $\lambda_2 = -\sqrt{\frac{4 \Lambda}{3}}$, so it is 
a stable node. 
The eigenvalues associated with $(\tilde{\Phi},0)$ have opposite sign, 
$\lambda_1 = \sqrt{\frac{\Lambda}{3}}$ and $\lambda_2= -2\sqrt{\frac{4 \Lambda}{3}}$, 
hence it is a saddle point.  
So the solution $\Phi\to \tilde{\Phi}$ is ``unstable'' and only with an exact initial 
value $\Phi_0$ can one reach the fixed point $(\Phi_\infty,\dot{\Phi}_\infty)=(\tilde{\Phi},0)$. 
In contrast, once we find some $\Phi_0$ value which yields a solution tending to $0$, 
we can vary $\Phi_0$ in a neighbourhood of this value and still go to the fixed point $(0,0)$. 

By continuity, if we take $\Phi_0$ slightly larger than $\Phi_{0,min}$, $\Phi(t)$ 
will still go to infinity in a finite time.  
Further increasing $\Phi_0$, two possibilities arise:

a) For $\Phi_{0,min} \le \Phi_0 < \Phi_{0,cr}$, where $\Phi_{0,cr}$ is some critical value,
$\Phi(t)$ diverges in a finite time. 

b) For $\Phi_{0,cr} < \Phi_0 < \Phi_{max}$ solutions will tend to zero. 

In the absence of $\Phi_{0,cr}$ only the first behaviour holds. 
For the specific case $\Phi_0 =\Phi_{0,cr}$, the solution will tend to $\tilde{\Phi}$. 
So this solution which tends to the saddle point is the separatrix which separates 
the two types of behaviours.
Hence for given parameters $\kappa$, $\Lambda$ and $c$ satisfying \eqref{int2}, the 
two types of behaviours are possible provided the existence of the separatrix.

\begin{figure}
\begin{centering}
\includegraphics[scale=.7]{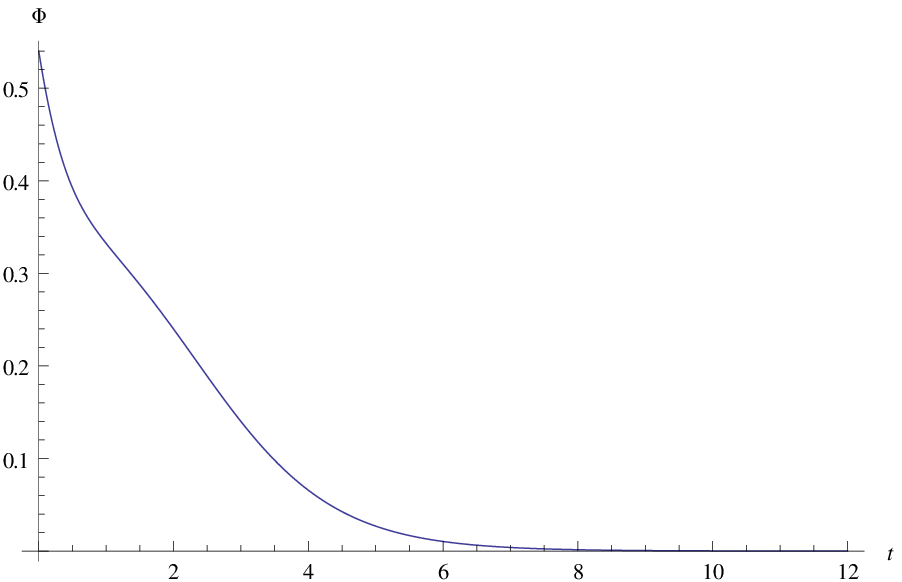}~~
\includegraphics[scale=.7]{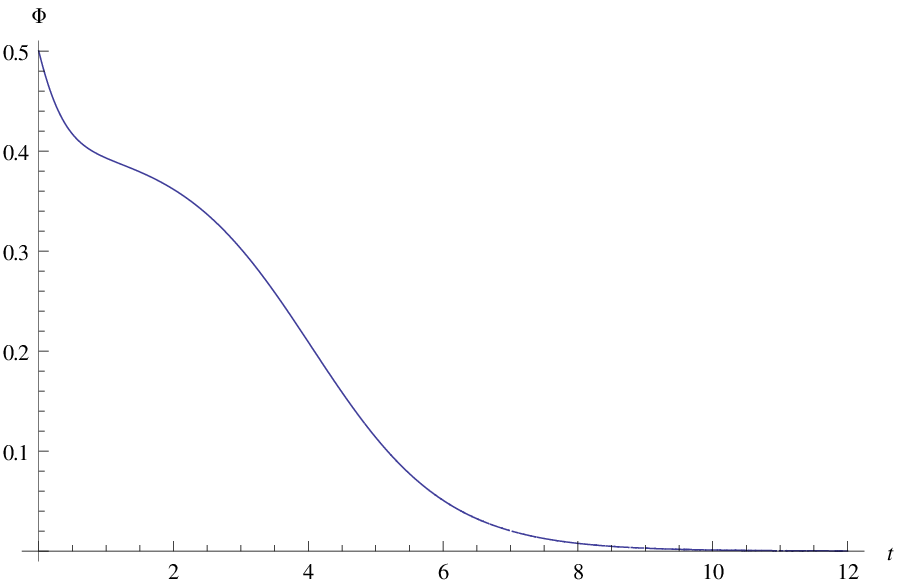}
\par\end{centering}
\caption{The function $\Phi(t)$ is shown starting at $t=0$ at the bounce for the same model 
parameters as in Figure 1. On the left panel, the initial value is $\Phi_0=0.54$ while on 
the right panel $\Phi_0=0.50$. As explained in the text $\ddot{\Phi}$ is positive above 
$\tilde\Phi$ whose value here is $\tilde\Phi = 0.40824829...$, but can change sign below 
$\tilde\Phi$. The curve on the right has a more pronounced feature because it starts 
closer to $\Phi_{0,cr} = 0.4936860...$. Both curves tend asymptotically to zero and 
constitute viable bouncing solutions. The curve starting \emph{precisely} at $\Phi_{0,cr}$ 
will tend asymptotically to $\tilde\Phi$.}
\label{fig3}
\end{figure}

Determination of the initial value $\Phi_{0,cr}$ yielding the separatrix is a 
well-known problem in the theory of ordinary differential equations, known 
as the connexion problem \cite{BFG07}. 
For given values of the parameters of the system, it is not possible in general to 
determine analytically whether $\Phi_{0,cr}$ exists, neither to find a closed form 
expression for it when it does exist.
We were able to find numerically $\Phi_{0,cr}$ for certain values of the 
parameters. For instance, for $\kappa^2 = 20$, $c = 3$ and $\Lambda = 3$, 
we find $\Phi_{0,cr} = 0.4936860...$. An example of viable solutions is shown 
on Figure \ref{fig3}. 

Using the analytical results given earlier, one can derive an analytical expression 
for $\Phi_{0,cr}$
\beq
\Phi_{0,cr} = - \left( \frac{\Lambda}{c \kappa^2}\right)^{\frac{1}{4}} 
\frac{1}{{\rm dn}\left( \left( \frac{2j+1}{\sqrt{2}} - 
       \left(\frac{9c}{\Lambda \kappa^2}\right)^{\frac{1}{4}} \right) 
       \frac{\Gamma^2\left(\frac{1}{4}\right)}{4\sqrt{\pi}}~|~2\right)}~.\label{Phi0cr} 
\eeq 
Let us explain how \eqref{Phi0cr} is obtained. As we look for the initial value $\Phi_{0,cr}$ 
for which $\Phi\to \tilde{\Phi}$, it is clear from \eqref{Phit} that $u(t=\infty)$ must then 
correspond to a zero of the function ${\rm dn}(u~|~2)$, which are given by
\beq
\frac{(2j+1)~\Gamma^2\left(\frac{1}{4}\right)}{4\sqrt{2\pi}} ,
                ~~~~~~~~~~~~~~~~~~~~~~~~~~j=0,1,...~.\label{j1}
\eeq
Using \eqref{etainfty}, we obtain
\beq
\frac{(2j+1)\Gamma^2\left(\frac{1}{4}\right)}{4\sqrt{2\pi}}=
{\rm dn}^{-1}
\left(-\left( \frac{\Lambda}{c \kappa^2}\right)^{\frac{1}{4}} \frac{1}{\Phi_{0,cr}}\,|\,2 \right)+
\left( \frac{c}{\Lambda\kappa^2} \right)^{\frac{1}{4}}\sqrt{3}\,
                          \frac{\Gamma^2\left(\frac{1}{4}\right)}{4\sqrt{\pi}}.\label{j2}
\eeq
Inverting this expression we obtain \eqref{Phi0cr}.
 
The value of $j$ must be chosen in such a way that $\Phi_{0,cr}$ lies inside the 
allowed interval \eqref{int2} and that \eqref{j2} is satisfied. As the function 
${\rm dn}(u|2)$ is periodic, only a small number of values of $j$ must be considered. 
The numerical value of $\Phi_{0,cr}$ given above was obtained from \eqref{Phi0cr} 
for $j=1$. For some values of the model parameters and any value of $j$, $\Phi_{0,cr}$ is 
not in the allowed interval and then the resulting model is not viable. 

In the presence of radiation one checks easily that the allowed range of initial values 
shrinks with a new larger minimal value $\Phi'_{0,min}$ satisfying 
\beq
\Phi'^4_{0,min} = \Phi^4_{0,min} \frac{A}{A'}~. 
\eeq
We have of course $A < A' < 0$.
The quantities $\tilde{\Phi},~\Phi_{max}$ on the other hand depend on the model parameters 
and remain unchanged. One can just repeat the proof of the existence of viable bouncing 
solutions. Hence, as long as we have $\Phi'_{0,min} < \Phi_{0,cr}$, the set of viable bouncing 
solutions remains. In the new expression for $\Phi(t)$ we will have 
$u \to {\rm dn}^{-1} \left( -\left( \frac{A}{A'}~\frac{\Lambda}{c \kappa^2}\right)^{\frac{1}{4}} 
  \frac{1}{\Phi_0}~|~2 \right) - i\sqrt{2}\left( \frac{A}{A'}~\frac{9c}{\kappa^2\Lambda} 
  \right)^{\frac{1}{4}}  F\left(i\sqrt{\frac{\Lambda}{3}}~t~|~2\right)$ while the same corrective 
factor $\left( \frac{A}{A'} \right)^{\frac{1}{4}}$ appears in front of \eqref{Phit} as well. We 
have the obvious corresponding change in \eqref{j2} and \eqref{Phi0cr}.  

Let us consider next the possibility to have a viable bounce when $Z=-1$ in 
the absence of radiation.  
If we want to keep the conformal invariance, the absence of ghosts 
($\omega_{BD}>-\frac{3}{2}$) and the existence of a bounce, we are led to the 
following ansatz:
\beq  
F_{Z=-1}(\Phi) = - F_{Z=1}(\Phi)~~~~~~~~~~~~~~~~~~U_{Z=-1}(\Phi) = - U_{Z=1}(\Phi)~,\label{Z-1} 
\eeq
with the parameters $\kappa^{-2},~\Lambda$ still positive. The quantities $F_{Z=1}$, resp. 
$U_{Z=1}$, correspond to \eqref{F}, resp. \eqref{U}. It is easily checked that 
\eqref{Eq1} is retrieved leading to a bounce provided the quantity $A$ is negative. 
This implies here too that $c>0$, but now $U_{Z=-1}$ no longer represents an inverted 
potential and we have $U_{Z=-1}(\Phi=0)<0$. 
As we have emphasized earlier, with \eqref{Z-1} the same equation of motion for the field 
$\Phi$ is obtained therefore leading to the same solution \eqref{Phit}.
However, the crucial difference is that the domain of physical validity differs 
in both cases: now all values $\Phi > \frac{\sqrt{6}}{\kappa}$ satisfy $F_{Z=-1}>0$. 
At the bounce, it is seen from \eqref{Fr1} that the potential must satisfy 
$U_{Z=-1}(\Phi_0)\ge 0$ yielding in this case too $\Phi_0 \ge \Phi_{0,min}$.
 
Up to these differences, we can essentially repeat the analysis performed for $Z=1$. 
It is easy to show 
using \eqref{ddPhi} that $\dot{\Phi}=0$ implies $\ddot{\Phi} > 0$ for $\Phi > \tilde{\Phi}$, 
and $\ddot{\Phi} < 0$ for $\Phi < \tilde{\Phi}$. On the other hand from \eqref{Fr2}, 
$\dot{\Phi}=0$ implies $\ddot{\Phi} < 0$. Hence $\dot{\Phi}=0$ is not possible for 
$\Phi > \tilde{\Phi}$ while it would correspond to a maximum for $\Phi < \tilde{\Phi}$.
On the other hand, from \eqref{ddPhi} only $\tilde{\Phi}$ remains an acceptable finite 
limit of $\Phi$ because $\Phi=0$ is now excluded. Imposing the condition 
$\frac{\Lambda \kappa^2}{36 c} > 1$, we now obtain 
\beq
\frac{\sqrt{6}}{\kappa} < \Phi_{0,min} < \tilde{\Phi}~.\label{Int}
\eeq  
If $\Phi_0 > \tilde{\Phi}$, $\dot{\Phi}_0 > 0$ implies $\Phi\to \infty$ while 
$\dot{\Phi}_0 < 0$ can in principle produce two possible behaviours namely 
$\Phi\to \tilde{\Phi}$ (in that case $\Phi_0=\Phi_{0,cr}$) or $\Phi\to 0$, the latter 
being unviable. The first possibility cannot occur. Indeed with $\dot{\Phi}_0 < 0$, some 
initial values above as well as below $\Phi_{0,cr}$ would tend to zero. But the particular 
value $\Phi_{0,cr}$ must separate two different behaviours, hence this is not possible. 
When $\Phi_0 < \tilde{\Phi}$, $\Phi\to 0$ for a negative initial slope because there is 
no minimum. With $\dot{\Phi}_0 > 0$, three different behaviours can be found: $\Phi\to 0$ 
when $\Phi$ goes through a maximum (below $\tilde{\Phi}$), $\Phi\to \infty$ when there is 
no maximum and finally  $\Phi\to \tilde{\Phi}$ for $\Phi_{0,cr}$ which now separates the 
two behaviours mentioned earlier, its existence is therefore allowed in this case. 
It is interesting that while a viable bounce can exist for $Z=-1$ with a potential bounded 
from below, it has measure zero in the initial conditions, only one particular set of initial 
conditions at the bounce is acceptable. The expression of $\Phi_{0,cr}$ for this case is 
also given by \eqref{Phi0cr}.

\section{Conclusions}

We have shown that for $Z=1$ only three types of solutions are possible. All stable 
solutions vanishing asymptotically are especially interesting, their initial value 
$\Phi_0$ at the bounce can vary in some range 
$\Phi_{0,cr} < \Phi_0 < \sqrt{6}\kappa^{-1}$.
The solution with $\Phi_0 = \Phi_{0,cr}$ is unique (measure zero) and moreover unstable. 
The stable solutions lead to the General Relativity limit if we assign to the bare 
gravitational constant $\kappa^2$ its numerical value in GR. 
Consistency of our model requires then the inequality $\frac{\Lambda~\kappa^2}{8 \pi}\ll 1$. 
If we want to embed our model in a realistic cosmology, this attractive feature is even 
reinforced by the possibility to add a radiation component to our bouncing model.  

Further, due to the invariance of the equations under the change $t \rightarrow -t$, 
the solutions for $t<0$ are obtained from those found for $t>0$ changing the sign of 
$\dot{\Phi}$. 
Hence, regular solutions for $t>0$ will diverge in the past both for $Z=1$ and $Z=-1$. 
Still, starting with regular initial values at $-\infty < t < 0$, it is possible to have 
a bouncing universe regular in the future. 
Note that a solution symmetric around the bounce requires ${\dot \Phi}_0=0$ which will 
diverge both in the past and in the future.

It is well known that a bounce in the early universe can cure many of the problems occurring 
in Big Bang cosmology besides of course avoiding the Big Bang singularity itself. Without 
exploring further this scenario, we have shown that surprisingly such a bounce can occur 
for a FLRW flat universe within one of the oldest and best understood alternatives to GR 
in a model tending dynamically to GR after the bounce.  The key point is to admit a negative
scalar field potential that can lead to a transient negative energy density of the scalar 
field for some range of parameters.\footnote{In this connection, the early paper~\cite{RR70} 
has to be mentioned, too, where a FLRW bounce in $f(R)$ gravity was achieved in case of the 
scalar degree of freedom in this theory (scalaron) being tachyonic permanently that 
corresponds to a negative (and even unbounded from below) effective scalar field potential 
in the Einstein frame, see also~\cite{IT12}.}  Of course, to make the model cosmologically 
viable, one has to generalize it somehow to open channels for decay of the effective 
positive (and large) cosmological constant $\Lambda$ into other quantum fields including 
those of the Standard Model of elementary particles at late times.

\section*{Acknowledgments}
A.S. was partially supported by the RFBR Grant No. 14-02-00894 and by the Russian Government 
Program of Competitive Growth of the Kazan Federal University.

\end{document}